\documentclass[12pt]{iopart}
\usepackage{iopams}
\expandafter\let\csname equation*\endcsname\relax
\expandafter\let\csname endequation*\endcsname\relax
\usepackage{amsmath}

\usepackage{graphicx}       
\graphicspath{{Figures/}} 

\def\nm{{\ {\rm nm}}}						

\def\Hz{{\, {\rm Hz}}}						
\def\kHz{{\, {\rm kHz}}}						
\def\MHz{{\, {\rm MHz}}}						

\def\ms{{\, {\rm ms}}}						

\def\Rb87{^{87}\rm{Rb}}					

\def\ex{{\mathbf e}_x}                            
\def\ey{{\mathbf e}_y}                            
\def\ez{{\mathbf e}_z}                            
\DeclareMathAlphabet\mathbfcal{OMS}{cmsy}{b}{n}
\def\El{E_{\mathrm{L}}}
\def\kl{k_{\mathrm{L}}}

\newcommand{\ket}[1]{|#1\rangle}
\newcommand{\braket}[2]{\langle #1|#2\rangle}

\begin{document}
		
\title{Fourier transform spectroscopy of a spin-orbit coupled Bose gas}
	
\author{A. Vald\'es-Curiel, D. Trypogeorgos, E. E. Marshall, I. B. Spielman}
\address{Joint Quantum Institute, University of Maryland and National Institute of Standards and Technology, College Park, Maryland, 20742, USA}
\date{\today}

\begin{abstract}


	We describe a Fourier transform spectroscopy technique for directly measuring band structures, and apply it to a spin-1 spin-orbit coupled Bose-Einstein condensate. In our technique, we suddenly change the Hamiltonian of the system by adding a spin-orbit coupling interaction and measure populations in different spin states during the subsequent unitary evolution. We then reconstruct the spin and momentum resolved spectrum from the peak frequencies of the Fourier transformed populations. In addition, by periodically modulating the Hamiltonian, we tune the spin-orbit coupling strength and use our spectroscopy technique to probe the resulting dispersion relation. The frequency resolution of our method is limited only by the coherent evolution timescale of the Hamiltonian and can otherwise be applied to any system, for example, to measure the band structure of atoms in optical lattice potentials.

\end{abstract}

\maketitle

\section*{Introduction}


Cold-atom systems offer the possibility of engineering single-particle dispersions that are analogues to those present in condensed matter systems, thereby creating exotic atomic `materials', with interaction-dominated or topologically non-trivial band structures \cite{lindner_floquet_2011,radic_strong_2015}. The properties of such materials depend on their underlying band structure, and a multitude of techniques have been developed for measuring the single particle dispersion relation. Here we present a Fourier transform technique that employs the connection between the energy spectrum of a system and its dynamics. This connection has been exploited to study the spectrum of both condensed matter \cite{jonas_two-dimensional_2003} and cold atom systems \cite{yoshimura_diabatic-ramping_2014,wang_atom-interferometric_2015} alike. Here we implemented a Fourier transform spectroscopy technique and applied it to spin-orbit coupled (SOC) Bose-Einstein condensates (BECs) to obtain their dispersion relation.

Spin-orbit coupling, naturally present in two-dimensional electron systems subject to an electric field perpendicular to the plane, is a necessary ingredient for phenomena such as the spin quantum Hall effect, and plays an important role in topological materials \cite{bychkov_oscillatory_1984-1,hasan_textitcolloquium_2010}. We engineered a Hamiltonian that has equal contributions of Rashba and Dresselhaus SOC \cite{lin_spin-orbit-coupled_2011}, in an ultra-cold atomic system by coupling the internal degrees of freedom of $^{87}\mathrm{Rb}$ atoms using  two laser fields \cite{dalibard_textitcolloquium_2011}. The fields change the spin state while imparting momentum to the system via two-photon Raman transitions \cite{lan_raman-dressed_2014,campbell_magnetic_2016}. The SOC term in the Hamiltonian can be made tunable by adding a periodic amplitude modulation in the Raman field \cite{jimenez-garcia_tunable_2015}.

Unlike the previous techniques used to measure the SOC dispersion in atomic systems \cite{cheuk_spin-injection_2012}, ours relies only on the unitary evolution of an initial state suddenly subjected to a SOC Hamiltonian and measuring occupation probabilities in a basis that does not diagonalize the Hamiltonian. In general, the initial state is not an eigenstate of the spin-orbit coupled Hamiltonian and undergoes unitary evolution. The spectral components of this time evolution are given by their relative energies, and using this time-domain evolution as a spectroscopic tool is useful for studying the energy spectrum of more complex time-dependent periodically driven systems \cite{jimenez-garcia_tunable_2015,eckardt_superfluid-insulator_2005,goldman_periodically_2014}, which are well suited for engineering and tuning Hamiltonians. 

This article is organized as follows. First we give a general description of the Fourier transform spectroscopy technique. We then describe the experimental procedure used to generate the spin-orbit coupling interaction in $^{87}$Rb BECs and apply the Fourier spectroscopy technique. Lastly we show the relative energies of our system and recover the SOC spectrum using the effective mass of the ground state. 

\subsection*{Operating principle of Fourier spectroscopy}

We focus on a system where we can measure the occupation probabilities of a set of orthonormal states $\{\ket{\psi_i}\}$ that fully span the accessible Hilbert space of the system. We then consider the time evolution of an arbitrary initial state $\ket{\Psi_0}=\sum\limits_{i}a_i\ket{\psi_i}$ as governed by a Hamiltonian $\hat{H}'(\{\Omega_i \})$ and observe the occupation probabilities of the $\{\ket{\psi_i}\}$ states of the measurement basis as a function of time. When $\hat{H}'$ is applied, the evolution of the initial state is $\ket{\Psi(t)}=\sum\limits_{i,j}a_ic_{i,j}e^{-iE'_jt/\hbar}\ket{\psi'_j}$, where $E'_j$ and $\ket{\psi'_j}$ are the eigenenergies and eigenstates of $\hat{H}'$, and $c_{i,j}(t)=\braket{\psi_i}{\psi'_j}$. The probability  
\begin{align}
P_k(t)=&|\braket{\psi_k}{\Psi(t)}|^2=\lvert\sum\limits_{i,j}a_ic_{i,j}c^{*}_{j,k}e^{-iE'_jt/\hbar}\rvert^2
\label{Eq:Probability}
\end{align}
of finding the system in a state $\ket{\psi_k}$ of the measurement basis can be expressed as a sum of oscillatory components, with amplitude given by the magnitude of the overlap integrals 
\begin{align}
P_k(t)=1+\sum\limits_{i,j\neq l} 2\lvert a_ic_{i,j}c_{j,k}c_{i,j'}c_{k',k}\rvert \cos(2\pi f_{jj'}t),
\end{align}
where $f_{jj'}=(E'_{j}-E'_{j'})/h$ is the frequency associated with the energy difference of two eigenstates of the Hamiltonian.
Fourier spectroscopy relies on measuring the occupation probabilities of each state in the measurement basis as a function of time, and extracting the different frequency components $f_{jj'}$ directly by computing the discrete Fourier transform. The bandwidth and frequency resolution of the measurement are determined by the total sampling time and the number of samples. For $N$ samples separated by a time interval $\Delta t$, the highest measured frequency will be $f_{\mathrm{bw}}=1/2\Delta t$, with resolution $\Delta f=1/\Delta tN$.

\begin{figure*}[bt]
	\begin{center}
		\includegraphics{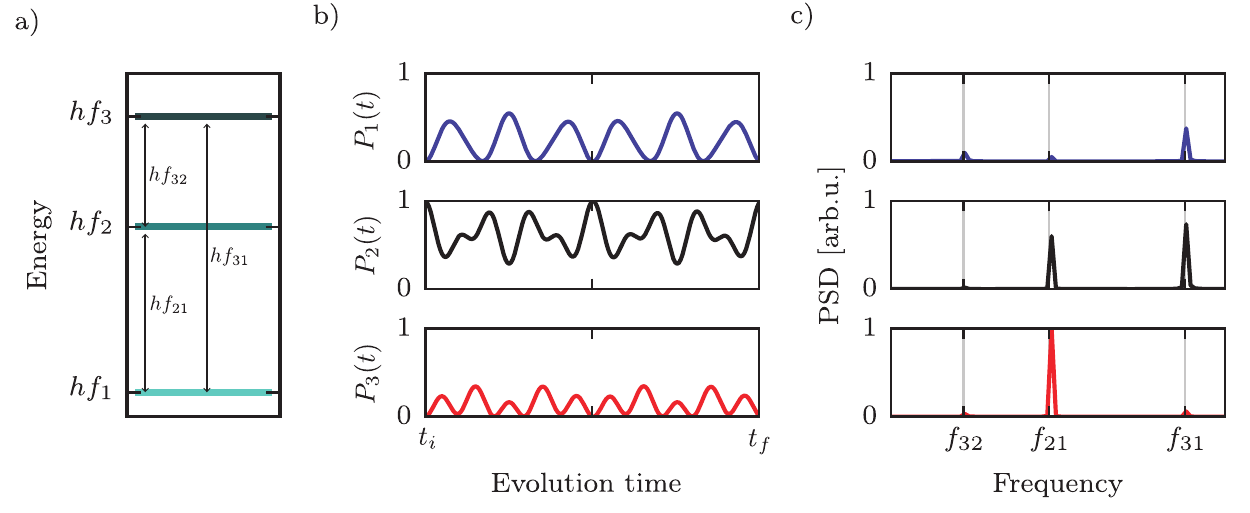}
		\caption
		{
			{\bf a)} Eigenenergies of a three-level system described by $\hat{H}'(\Omega_1,\Omega_2,\Omega_3)$. 
			{\bf b)} The system is prepared in $\ket{\psi_2}$ and subjected to $\hat{H}'$ at time $t_i$. The three panels show the occupation probabilities of the states $\ket{\psi_1}$ (blue), $\ket{\psi_2}$ (black), and $\ket{\psi_3}$ (red) in the measurement basis, for evolution times up to $t_f$. 
			{\bf c)} Power spectral density of the occupation probabilities from b). The three peaks in the Fourier spectra correspond to the energy differences present in a).
		\label{fig:Figure1}}
	\end{center}
\end{figure*}

Figure~\ref{fig:Figure1} illustrates the principle of Fourier spectroscopy for a three level system, initially prepared in the state $\ket{\Psi_0}=\ket{\psi_2}$, subject to the Hamiltonian
\begin{equation}
\hat{H}'=\begin{pmatrix}
E_1 & 0 & 0  \\
0 & E_2 & 0  \\
0 & 0 & E_3 \\
\end{pmatrix}
+\begin{pmatrix}
0 & \Omega_1 & \Omega_2  \\
\Omega_1^{*} & 0 & \Omega_3  \\
\Omega_2^{*} & \Omega_3^{*} & 0
\end{pmatrix},
\end{equation}
where we measure the occupation probability as a function of time for each of the $\{\ket{\psi_1}, \ket{\psi_2},\ket{\psi_3}\}$ states. The three eigenenergies $E'_i=hf_i$ are displayed in figure~\ref{fig:Figure1}(a). The three energy differences $hf_{jj'}$ between the levels determine the oscillation frequencies of the occupation probabilities, as can be seen in figure~\ref{fig:Figure1}(b). Finally, a plot of the power spectral density (PSD) in figure~\ref{fig:Figure1}(c) shows three peaks at frequencies corresponding to the three relative energies of $\hat{H}'$.

\section*{Experiment}

We begin our experiments with a $^{87}$Rb BEC \cite{lin_rapid_2009} containing about $4\times 10^4$ atoms in the $5^2S_{1/2}$ electronic ground state, and in the $\ket{f=1,m_f=-1}$ hyperfine state. The BEC is confined in a crossed optical dipole trap formed by two $1064\nm$ beams propagating along $\ex+\ey$ and $\ex-\ey$, which give trapping frequencies   $(\omega_x,\omega_y,\omega_z)/2\pi=(42(3),34(2),133(3))\Hz$\footnote{All uncertainties herein represent the uncorrelated combination of statistical and systematic errors.}. We break the degeneracy of the three $m_F$ magnetic sub-levels by applying a $1.9893(3)\,$mT bias field along $\mathbf{e}_z$ that produces a $\omega_Z / 2\pi  = 14.000(2) \MHz$ Zeeman splitting, and a quadratic Zeeman shift $\epsilon$ that shifts the energy of $\ket{f=1,m_F=0}$ by $-h\times 28.45\, \kHz$. We adiabatically transfer our BEC into $\ket{f=1,m_F=0}$ by slowly ramping the bias field, from  $B_i=1.9522(3)\,$mT to $B_f=1.9893(3)\,$mT in $50\,$ms while applying a $14\,\MHz$ radio-frequency magnetic field with approximately $20\,$kHz coupling strength that was ramped on $50$ ms before the bias field. We then apply a pair of $250\,\mu\mathrm{s}$ microwave  pulses that each transfer a small fraction of atoms into the $5^2{\rm S}_{1/2}$ $f=2$ manifold that we use to monitor and stabilize the bias field \cite{l._j._leblanc_direct_2013}. The microwave pulses are detuned by $\pm 2\, \kHz$ from the $\ket{f=1,m_F=0}\leftrightarrow\ket{f=2,m_F=1}$ transition and spaced in time by $33\, \mathrm{ms}$ (two periods of $60\, \mathrm{Hz}$). We imaged the transferred atoms following each pulse using absorption imaging\footnote{We did not apply repump light during this imaging, so the untransferred atoms in the $f=1$ manifold were largely undisturbed by the imaging process.}, and count the total number of atoms $n_1$ and $n_2$ transferred by each pulse. The imbalance in these atom numbers $(n_1-n_2)/(n_1+n_2)$ leads to a $4\kHz$ wide error signal that we use both to monitor the magnetic field before each spectroscopy measurement and cancel longterm drifts in the field. 

\begin{figure*}[bt]
	\begin{center}
		\includegraphics{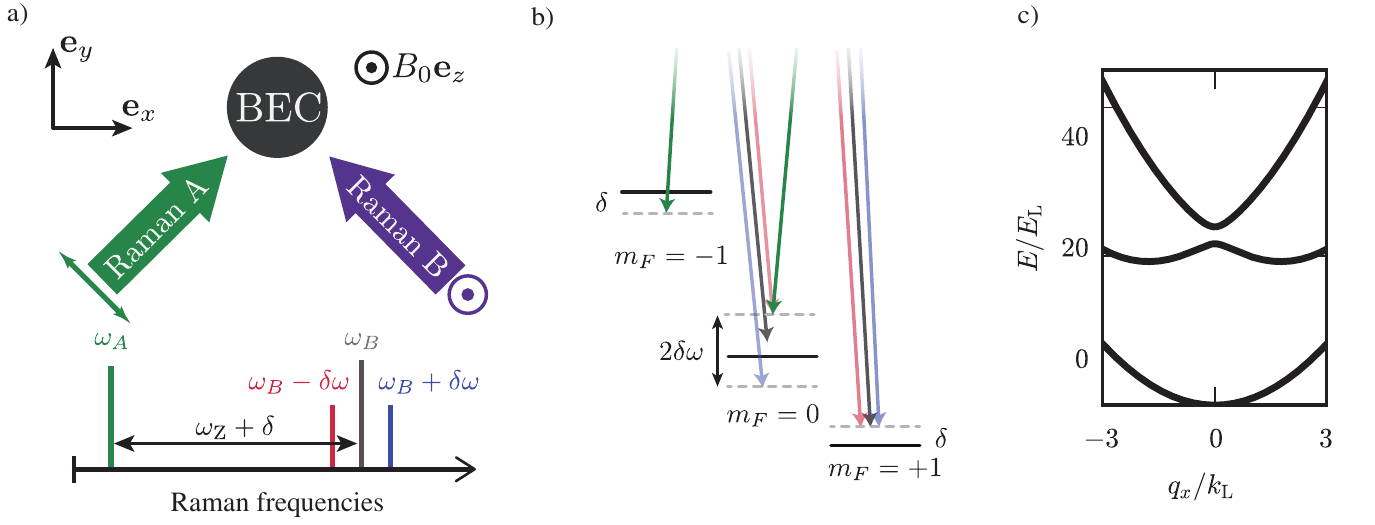}
		\caption
		{
			{\bf a)} Setup. A bias magnetic field $B_0\ez$, with $B_0=1.9893$\,mT splits the hyperfine energy levels of the $f=1$ manifold of $^{87}$Rb by $\omega_Z/2\pi=14$\,MHz. A pair of cross polarized Raman beams propagating along $\ex+\ey$ and $\ex-\ey$ couple the atoms' momentum and spin states. 
			{\bf b)} The Raman frequencies are set to $\omega_A=\omega_L+\delta$ and $\omega_B=\omega_L+\omega_Z$. We add frequency sidebands to $\omega_B$, separated by $\pm \delta\omega$. The amplitude modulation from the interference between the multiple frequency components results in tunable SOC.
			{\bf c)} SOC dispersion for Raman coupling strength $\Omega_0=12E_{\mathrm{L}}$ and $\Omega=0$, on four photon resonance.
		\label{fig:Figure2}}
	\end{center}
\end{figure*}

We induce spin-orbit coupling using a pair of intersecting, cross polarized `Raman' laser beams propagating along $\mathbf{e}_x+\mathbf{e}_y$ and $\mathbf{e}_x-\mathbf{e}_y$, as shown in figure~\ref{fig:Figure2}(a) and (b). This beams have angular frequency $\omega_A=\omega_L+\delta$ and $\omega_B=\omega_L+\omega_Z$, where $2\delta$ is the, experimentally controllable, detuning from four photon resonance between $m_F=-1$ and $m_F=+1$. The geometry and wavelength of the Raman fields determine the natural units of the system: the single photon recoil momentum $k_{\mathrm{L}}=\sqrt{2}\pi/\lambda_{\mathrm{R}}$ and its associated recoil energy $E_{\mathrm{L}}=\hbar^2k_{\mathrm{L}}^2/2m$, as well as the direction of the recoil momentum $\mathbf{k}_{\mathrm{L}}=k_{\mathrm{L}}\ex$. The Raman wavelength is $\lambda_{\mathrm{R}}=790.032\,\nm$, so that the scalar light shift is zero. 

Our system is well described by the Hamiltonian including atom-light interaction along with the kinetic contribution
 
 \begin{align}
 \begin{split}
 \hat{H}_{\mathrm{SOC}} = &\frac{\hbar^2q_x^2}{2m} + \alpha q_x\hat{F}_z  + 4E_{\mathrm{L}}\hat{\mathbb{I}} + \Omega_{\mathrm{R}}\hat{F}_x  +(4E_{\mathrm{L}}-\epsilon)(\hat{F}_z^2-\hat{\mathbb{I}}) +\delta\hat{F}_z,
 \label{Eq:SOCone}
 \end{split}
 \end{align}
where $q$ is the quasimomentum, $\hat{F}_{x,y,z}$ are the spin-1 angular momentum matrices,  $\alpha=\hbar^2k_{\mathrm{L}}/m$ is the SOC strength, and $\Omega_{\mathrm{R}}$ is the Raman coupling strength, proportional to the Raman laser intensity. The Raman field couples  $\ket{m_F=0,\, q=q_x}$ to $\ket{m_F=\pm1,\, q=q_x\mp 2k_{\mathrm{L}}}$, generating a spin change of $\Delta m_F=\pm1$ and imparting a $\mp 2k_{\mathrm{L}}$ momentum. The eigenstates of $\hat{H}_{\mathrm{SOC}}$ are linear combinations of these states and $\ket{m_F=0,\,q=q_x}$, and the set $\{\ket{m_F,q}\}$ constitutes the measurement basis for the Fourier transform spectroscopy.

Figure \ref{fig:Figure2}(c) shows a typical band structure of our spin-1 SOC system as a function of quasimomentum for a large and negative quadratic Zeeman shift $-\epsilon>4E_{\mathrm{L}}$. In this parameter regime the ground state band has a nearly harmonic dispersion with an effective mass $m^{*} = \hbar^2[d^2E(k_x)/d^2x]^{-1}$, only slightly different from that of a free atom. 

We engineer a highly tunable dispersion relation in which we can independently control the size of the gap at $q_x=0$ as well as the SOC strength $\alpha$ by adding frequency sidebands to one of the Raman beams. The state of the system can change from $\ket{m_F=-1,\,q=q_x+2k_{\mathrm{L}}}$ to  $\ket{m_F=1,\,q=q_x-2k_{\mathrm{L}}}$ by absorbing a red detuned photon first followed by a blue detuned photon and vice versa, in a similar way to the M\o lmer-S\o rensen entangling gate in trapped ion systems \cite{sorensen_entanglement_2000}. The interference of the multiple frequency components leads to an amplitude modulated Raman field giving an effective Floquet Hamiltonian with tunable SOC \cite{jimenez-garcia_tunable_2015}. When we set the angular frequencies of the sidebands to $\omega=\omega_{\mathrm{A}}+\omega_Z \pm \delta\omega$, the Hamiltonian (equation~\ref{Eq:SOCone}) acquires a time-dependent coupling $\Omega_{\mathrm{R}}(t)=\Omega_0 + \Omega\cos(\delta\omega t)$. This periodically driven system is well described by Floquet theory \cite{floquet_sur_1883}, and we calculate the spectrum of Floquet quasi-energies that are grouped into manifolds separated in energy by integer multiples of $\hbar\delta\omega$ as shown in figure~\ref{fig:Floquet}. We define an effective, time-independent Hamiltonian $\hat{H}_{Fl}$ that describes the evolution of the system sampled stroboscopically at an integer number of driving periods, with the time evolution operator $\hat{U}(t_0,t_0+T)=e^{-iT\hat{H}_{Fl}}$. For $\hbar\delta\omega \gg 4E_{\mathrm{L}}$, the Floquet Hamiltonian retains the form of equation~\ref{Eq:SOCone} with renormalized coefficients and an additional coupling term:

\begin{figure*}[bt]
	\begin{center}
		\includegraphics{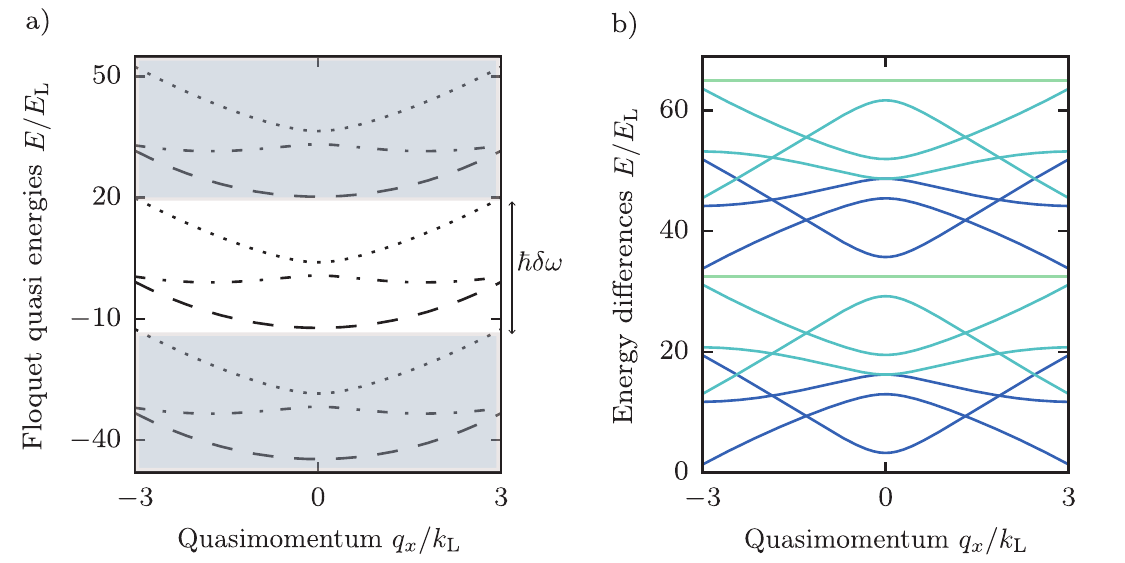}
		\caption
		{
			{\bf a)} Floquet quasi-energies of a three level Hamiltonian with SOC and time periodic coupling strength. The quasi-energies are grouped into manifolds consisting of three levels that get repeated with a periodicity equal to $\hbar\delta\omega$.
			{\bf b)} Energy differences of the Floquet quasi-energies. Each color represents the energy difference, separated by a fixed number of neighboring levels. When the number of neighboring levels is a multiple of 3, the energy differences are straight lines, a result of the periodic structure of the Floquet manifolds. 
		\label{fig:Floquet}}
	\end{center}
\end{figure*}

\begin{align}
\begin{split}
\hat{H}_{Fl} = &\hat{H}_{SOC}(q,\Omega_0,\tilde{\alpha},\tilde{\delta},\tilde{\epsilon}) + \tilde{\Omega}\hat{F}_{xz},
\label{Eq:SOCeff}
\end{split}
\end{align}
where $\tilde{\alpha}= J_0(\Omega/\delta\omega)\alpha$, $\tilde{\Omega}=1/4(\epsilon+4E_{\mathrm{L}}) [J_0(2\Omega/\delta\omega)-1]$, $\tilde{\delta}=J_0(\Omega/\delta\omega)\delta$, and $\tilde{\epsilon}= 1/4(4E_{\mathrm{L}}-\epsilon) -
1/4(4E_{\mathrm{L}} + 3 \epsilon) J_0( 2\Omega/\delta\omega)$.  $J_0$ is the the zero order Bessel function of the first kind, and $\hat{F}_{xz}$ is the $\hat{\lambda}_4$ Gell-Mann matrix that directly couples $\ket{m_f=-1, q=q_x+2k_{\mathrm{L}}}$ and $\ket{m_f=+1, q=q_x-2k_{\mathrm{L}}}$ states. The experimentally tunable parameters $\delta\omega$, $\Omega$ and $\Omega_0$ can be used to tune the SOC dispersion.

We use Fourier transform spectroscopy to measure the spectrum of the SOC Hamiltonian (equation~\ref{Eq:SOCeff}) for three coupling regimes: (i) $\Omega_0\neq0$ and $\Omega=0$, (ii)  $\Omega_0=0$ and $\Omega\neq0$ and (iii) $\Omega_0\neq0$ and $\Omega\neq0$. We turned on the Raman laser non-adiabatically, in approximately $1	\,\mu\mathrm{s}$. We let the system evolve subject to $\hat{H}_{\mathrm{SOC}}$ for up to $900\, \mathrm{\mu s}$, and  then turn off the laser while releasing the atoms from the optical dipole trap. We can resolve individual spin components by applying a spin-dependent `Stern-Gerlach' force using a magnetic field gradient. We then image the atoms using absorption imaging after a $21\,\mathrm{ms}$ time of flight. Our images reveal the atoms' spin and momentum distribution, allowing us to measure the fraction of atoms in each state of the measurement basis, and thereby obtain the occupation probability. The density of sampling points and the maximum evolution time are chosen so that the bandwidth of the Fourier transform is comparable to, or larger than, the highest frequency in the evolution of the system while maximizing resolution. Experimental decoherence is an additional constraint which becomes significant around 1 ms. 

In order to map the full spin and momentum dependent band structure of $\hat{H}_{\mathrm{SOC}}$, we measure the time dependent occupation probabilities at a fixed Raman coupling strength and different values of Raman detuning $\delta$, for the same initial state $\ket{m_F=0, q_x=0}$. For $\hat{H}_{\mathrm{SOC}}$, momentum and detuning are equivalent up to a numerical factor, $\delta/E_{\mathrm{L}}=4q_x/k_{\mathrm{L}}$, since the detuning term $\delta\hat{F}_z$ and the momentum term $\alpha\hat{q}_x\hat{F}_z$ have the same effect in the relative energies. This relation follows from the Doppler shift of the light frequency experienced by atoms moving relative to a light source: a stationary BEC in the laboratory reference frame dressed by a detuned laser field is equivalent to a moving BEC and a resonant laser field.

We control the frequency and the detuning of the Raman beams using two acousto-optic modulators (AOMs), one of which is driven by up to three phase coherent frequencies. For each of the three coupling cases that we measured, we applied the Raman beams at detuning values within the interval $\pm 12 E_{\mathrm{L}}$ which corresponds to quasimomentum values $\pm 3k_{\mathrm{L}}$.

\subsection*{Effective mass}
We recover the full spectrum of the system, rather than the relative energies, by measuring the effective mass of the nearly quadratic lowest branch of the dispersion, giving us an energy reference that we then use to shift the measured frequencies in the PSD. We measure the effective mass of the Raman dressed atoms by adiabatically preparing the BEC in the lowest eigenstate and inducing dipole oscillations. The effective mass of the dressed atoms  is related to the bare mass $m$ and the bare and dressed trapping frequencies $\omega$ and $\omega^{*}$ by the ratio $m^{*}/m=(\omega/\omega^{*})^2$. We measured this ratio following~\cite{lin_synthetic_2011}; we start in  $\ket{m_F=0,\, k_x=0}$ state and adiabatically turn on the Raman laser in $10\,\mathrm{ms}$ while also ramping the detuning to $\delta\approx0.5\,\El$, shifting the minima in the ground state energy away from zero quasi-momentum. We then suddenly bring the field back to resonance, exciting the BEC's dipole mode in the optical dipole trap. We measured the bare state frequency by using the Raman beams to initially induce motion but subsequently turn them off in $1\ms$ and let the BEC oscillate. For this set of measurements, we adjusted our optical dipole trap to give new trapping frequencies $(\omega_x, \omega_y, \omega_z)/2\pi=(35.9(4), 32.5(3), 133(3))\,\Hz$, nominally symmetric in the plane defined by $\ex$ and $\ey$. The Raman beams co-propagate with the optical dipole trap beams; therefore, the primary axes of the dipole trap frequencies are at a $45^{\circ}$ angle with respect to the direction of $\mathbf{k}_{\mathrm{L}}$. 

\begin{figure*}[ht]
	\begin{center}
		\includegraphics{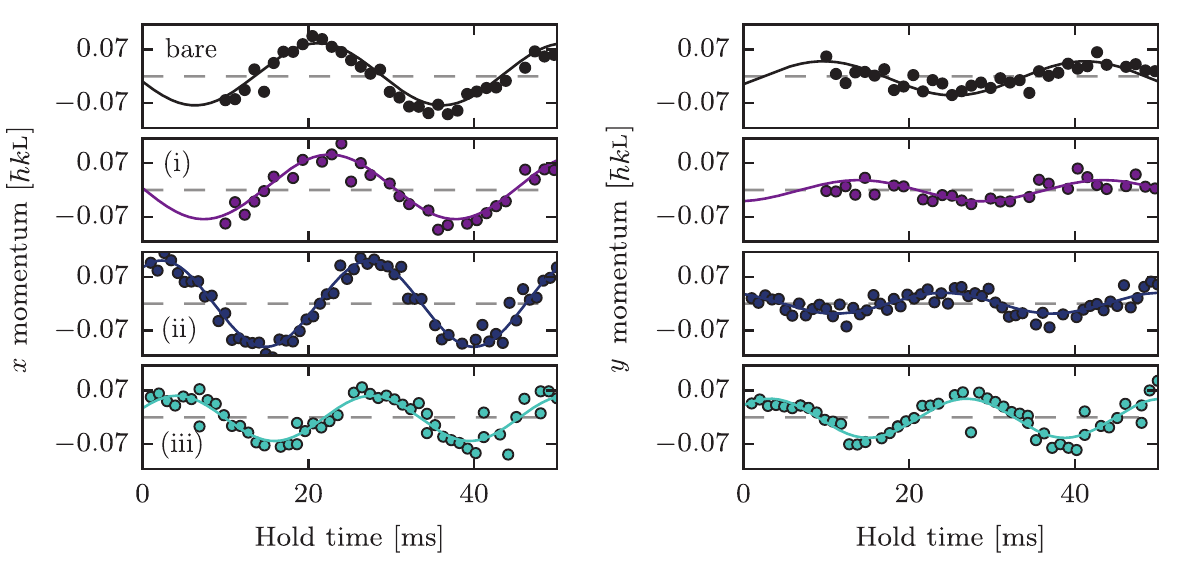}
		\caption
		{  Oscillation of the BEC in the dipole trap along the  recoil directions $\mathbf{e}_{x}$ and  $\mathbf{e}_{y}$ for (top) bare atoms, and the three parameter regimes that we explored (i), (ii), and (iii).  
		\label{fig:Figure4}}
	\end{center}
\end{figure*}

Figure~\ref{fig:Figure4} shows the dipole oscillations along the $\mathbf{e}_{x}$ and $\mathbf{e}_{y}$ directions for the three different coupling regimes we explored, as well as the bare state motion. The resulting mass ratios for the three coupling regimes are $m/m^{*}=$  (i) $1.04(8)$, (ii) $0.71(7)$, and (iii) $0.62(4)$.

\section*{Measured dispersion}

\begin{figure*}[t]
	\begin{center}
		\includegraphics{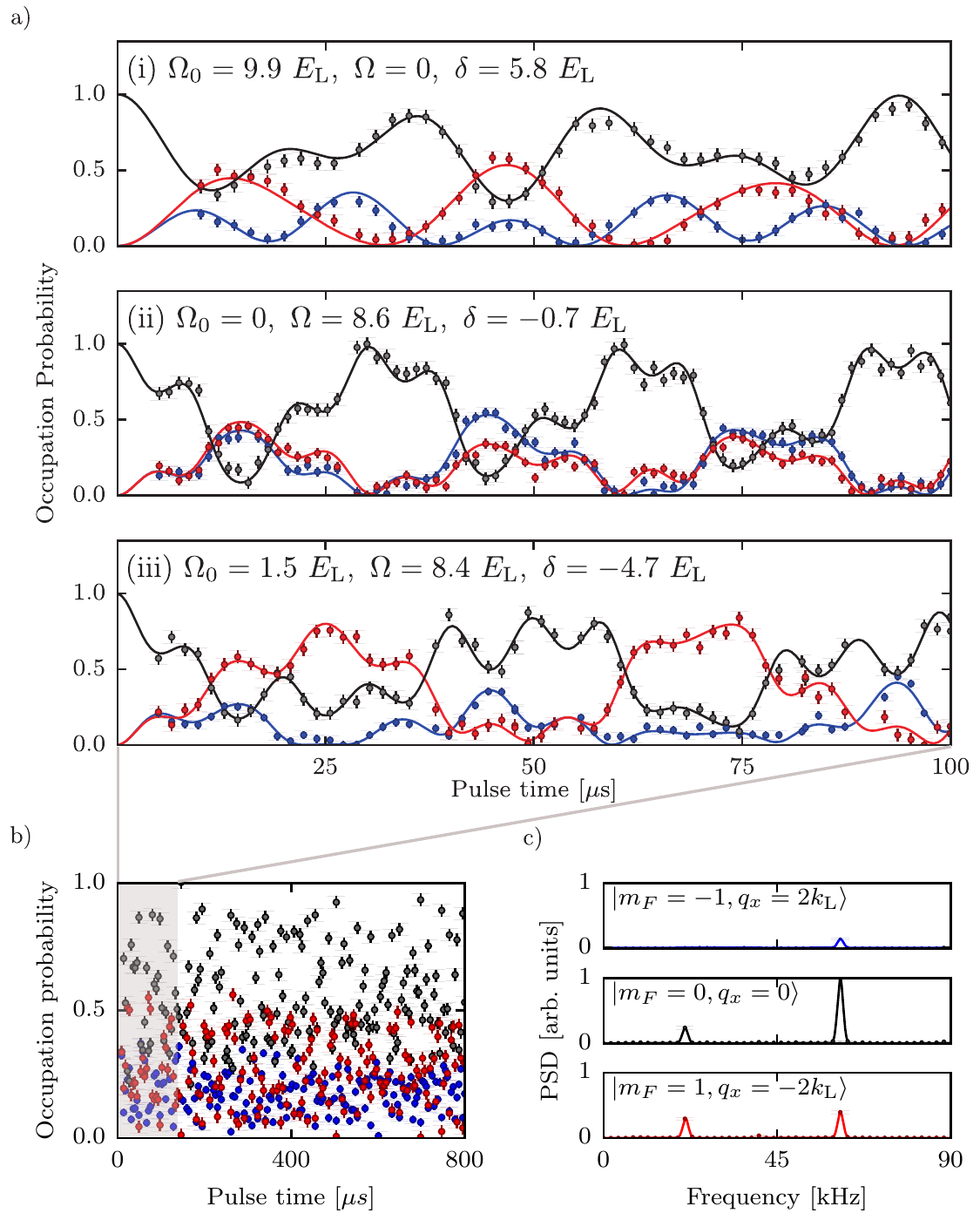}
		\caption
		{{\bf a)}
		Occupation probability for the three states in the measurement basis $\ket{m_f=-1, q=q_x+2\kl}$ (blue), $\ket{m_f=0,q=q_x}$ (black), and $\ket{m_f=+1,q=q_x-2\kl}$(red),  following unitary evolution under $\hat{H}_{\mathrm{SOC}}$ for times up to 100 $\mu s$ at different spin-orbit coupling regimes: (i) $\Omega_0=9.9 E_{\mathrm{L}}$, $\Omega=04$,  $\delta=5.8 E_{\mathrm{L}}$, (ii) $\Omega_0=0$, $\Omega=8.6\,E_{\mathrm{L}}$,  $\delta=-0.7\,E_{\mathrm{L}}$, $\delta\omega=\epsilon+12\,E_{\mathrm{L}}$, and (iii) $\Omega_0=1.5\,E_{\mathrm{L}}$, $\Omega=8.4\, E_{\mathrm{L}}$,  $\delta=-4.7\,E_{\mathrm{L}}$, $\delta\omega=\epsilon+17\,E_{\mathrm{L}}$.
		{\bf b)} Occupation probability for long pulsing up to 800 $\mu$s for parameters as in (iii). 
		{\bf c)} Power spectral density of the occupation probability. We subtract the mean value of each probability before taking the Fourier transform to remove peaks at $f=0$. The peaks in the PSD then correspond to the relative eigenenergies of $\hat{H}_{SOC}$.
		}
		\label{fig:Figure5}
	\end{center}
\end{figure*}

We mapped the band structure of spin-orbit coupled atoms for three different coupling regimes. Figure~\ref{fig:Figure5}(a) shows representative traces of the measured occupation probabilities for short evolution times along with fits to the unitary evolution given by $\hat{H}_{\mathrm{SOC}}$ with $\delta$, $\Omega_0$, and $\Omega$ as free parameters. The fit parameters agree well with independent microwave and Raman power calibrations. In the lower two panels, where the Raman coupling strength is periodically modulated, the occupation probabilities oscillate with more than three frequencies since the full description of the system is given by the Floquet quasi-energy spectrum. Figure~\ref{fig:Figure5}(b),(c) shows the occupation probabilities for the parameter regime (iii)  for longer evolution times along with the PSD of the occupation probability of each spin state. 

\begin{figure*}[t]
	\begin{center}
		\includegraphics{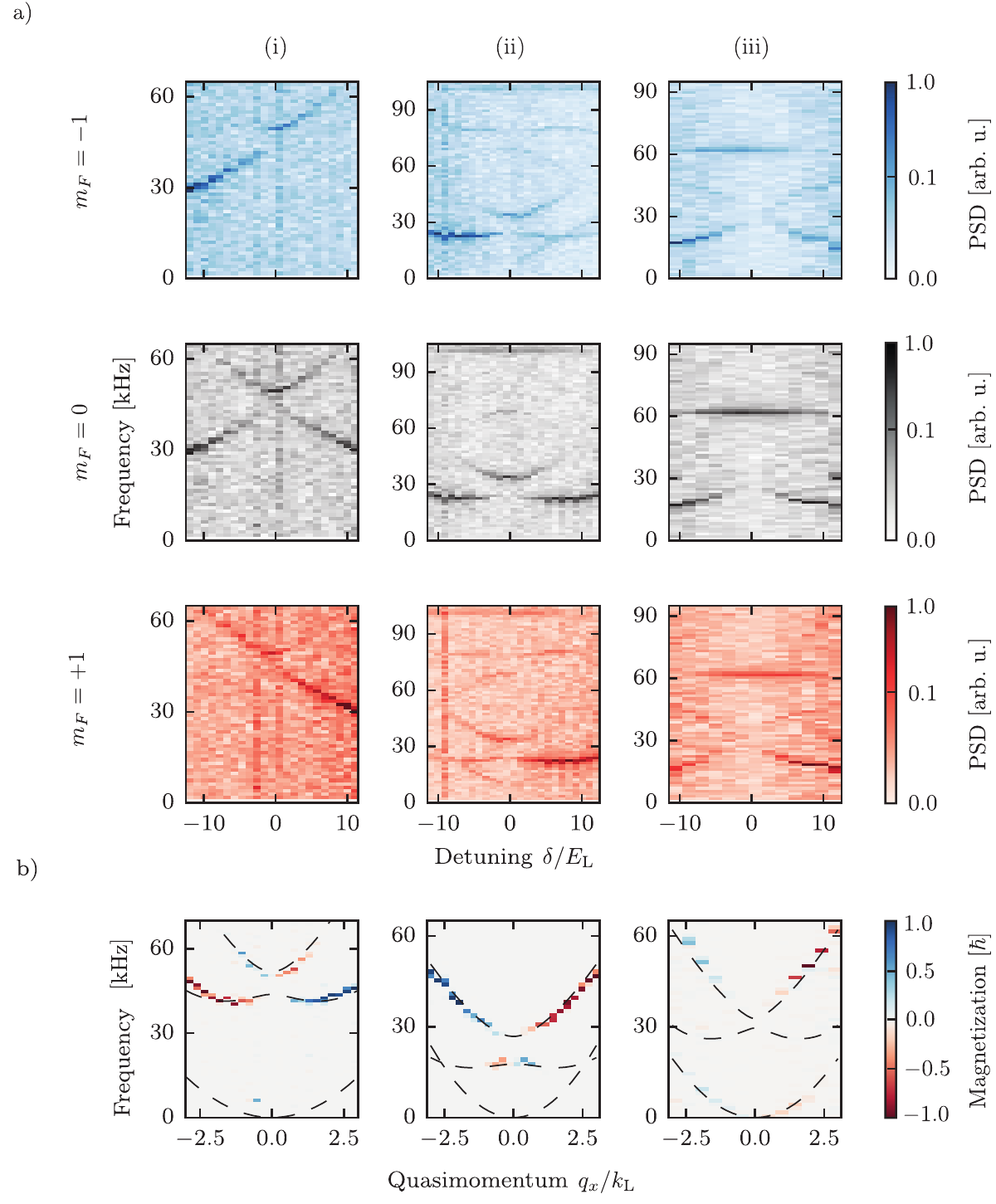}
		\caption
		{
			{\bf a)} Power spectral density of the time dependent occupation probability for each state in the measurement basis for three coupling regimes:
			{ (Left)} $\Omega_0=9.9 E_{\mathrm{L}}$, $\Omega=0$,
			{ (Center)} $\Omega_0=0$, $\Omega=8.6 E_{\mathrm{L}}$,  $\delta\omega=\epsilon+12 E_{\mathrm{L}}$, and
			{ (Right)} $\Omega_0=4.9 E_{\mathrm{L}}$, $\Omega=8.4 E_{\mathrm{L}}$,  $\delta\omega=\epsilon+17 E_{\mathrm{L}}$. Each panel is normalized to peak amplitude to highlight small amplitude features in the PSD of the periodically driven SOC, and the highest value on the frequency axis corresponds to the FFT bandwidth..
			{\bf b)} Spin-dependent SOC dispersion for three different coupling regimes. We combine the PSD of the occupation probability of the states $\ket{m_F=\pm 1, q_x=\mp 2k_{\mathrm{L}}}$, and shift each frequency by an amount proportional to the squared quasimomentum and the effective mass. The dashed lines are the calculated Floquet energies for the Hamiltonian using our calibration parameters. 
		}
		\label{fig:Figure6}
	\end{center}
\end{figure*}

We use a non-uniform fast Fourier transform algorithm (NUFFT) on a square window to obtain the power spectral density of the occupation probability since our data points are not always evenly spaced because of imperfect imaging shots. The heights of the peaks in the PSD are related to the magnitude of the overlap integrals between the initial state and the Raman dressed states. Figure~\ref{fig:Figure5}(c) shows the raw PSD of the time evolution of the system under $\hat{H}_{\mathrm{SOC}}$ for a given Raman coupling strength and detuning. We put together all the PSDs for the three coupling regimes in the spectra shown on the top three panels in figure~\ref{fig:Figure6}. Each column corresponds to a different coupling regime and the colors represent the different spin states of the measurement basis. The spectra show that some overlap integrals vanish near $\delta=0$, which is manifested as missing peaks in the PSD. The periodic structure of the Floquet quasi-energy spectrum gives rise to peaks at constant frequencies of $\delta\omega$ and $2\delta\omega$ independently of the Raman detuning, and a structure that is symmetric about the frequencies $2\pi f_1=\delta\omega/2$ and $2\pi f_2=\delta\omega$.

We obtain the characteristic dispersion of a SOC system after adding a quadratic term to the PSD, proportional to the measured effective mass,  and after rescaling the detuning into recoil momentum units. We combine the PSD of the time evolution of the three $\ket{m_F}$ states to look at the spin dependence of the spectra.  Figure \ref{fig:Figure6} shows the measured spectra as well as the Floquet quasi-energies calculated for the Hamiltonian parameters obtained from our calibrations. The spectral lines that can be resolved with our technique depend on the overlap integrals of the initial state with the target Hamiltonian eigenstates. Additional energies can be measured by repeating the experiment with different initial states. The spectral lines we were able to resolve are in good agreement with the calculated energies of the Hamiltonian.

\section*{Conclusion}

We measured the spin and momentum dependent dispersion relation of a spin-1 spin-orbit coupled BEC using a Fourier transform spectroscopy technique along with a measured effective mass of the initial state branch. We studied a periodically driven SOC system and found a rich Floquet quasi-energy spectrum. Our method can be applied generically to any system with long enough coherent evolution to resolve the energy scales of interest, and could prove particularly useful to study systems where it is harder to predict or compute the exact energies, such as cold atom realizations of disordered or highly correlated systems \cite{eisert_quantum_2015}. Moreover, this technique can be extended with the use of spectograms to study time dependent spectra, such as that of systems with quench-induced phase transitions.

\section*{Acknowledgements}
This work was partially supported by the ARO’s atomtronics MURI, the AFOSR’s Quantum Matter MURI, NIST, and the NSF through the PFC at the JQI. We are grateful for the careful reading if this manuscript by N. E. Lundblad, P. Solano, B. E. Anderson, L. A. Orozco and F. Salces-C\'arcoba.

\section*{Appendix A: Recovering the SOC dispersion from the PSD}
In this section we describe how we obtain a trapping frequency along an axis that is not defined by our optical dipole trap beams and how we use it to shift the PSD to obtain the absolute SOC spectrum from a spectrum of relative energies. 

The kinetic and potential terms in the Hamiltonian including the contribution of the Raman and optical dipole trap are

\begin{align}
\hat{H}_{\perp}= &\frac{\hbar^2q_x^2}{2m^{*}} + \frac{\hbar^2q_y^2}{2m}+\frac{m}{2}[\omega_{x'}^2x'^2+\omega_y'^2y'^2] \nonumber \\
= & \frac{\hbar^2}{2m^{\star}}k_x^2 + \frac{1}{2m}k_y^2+\frac{m}{2}[(\omega_{x'}^2+\omega_{y'}^2)(x^2+y^2)+2xy(\omega_{x'}^2-\omega_{y'}^2)],
\end{align}
where we have used  $x'=(x+y)/\sqrt{2}$ and   $y'=(x-y)/\sqrt{2}$. Therefore, for $\omega_{x'}=\omega_{y'}$, a simple rotation yields a trapping frequency along the Raman recoil direction $\omega_x^2=\omega_{x'}^2+\omega_{y'}^2$.

\begin{figure*}[ht]
	\begin{center}
		\includegraphics{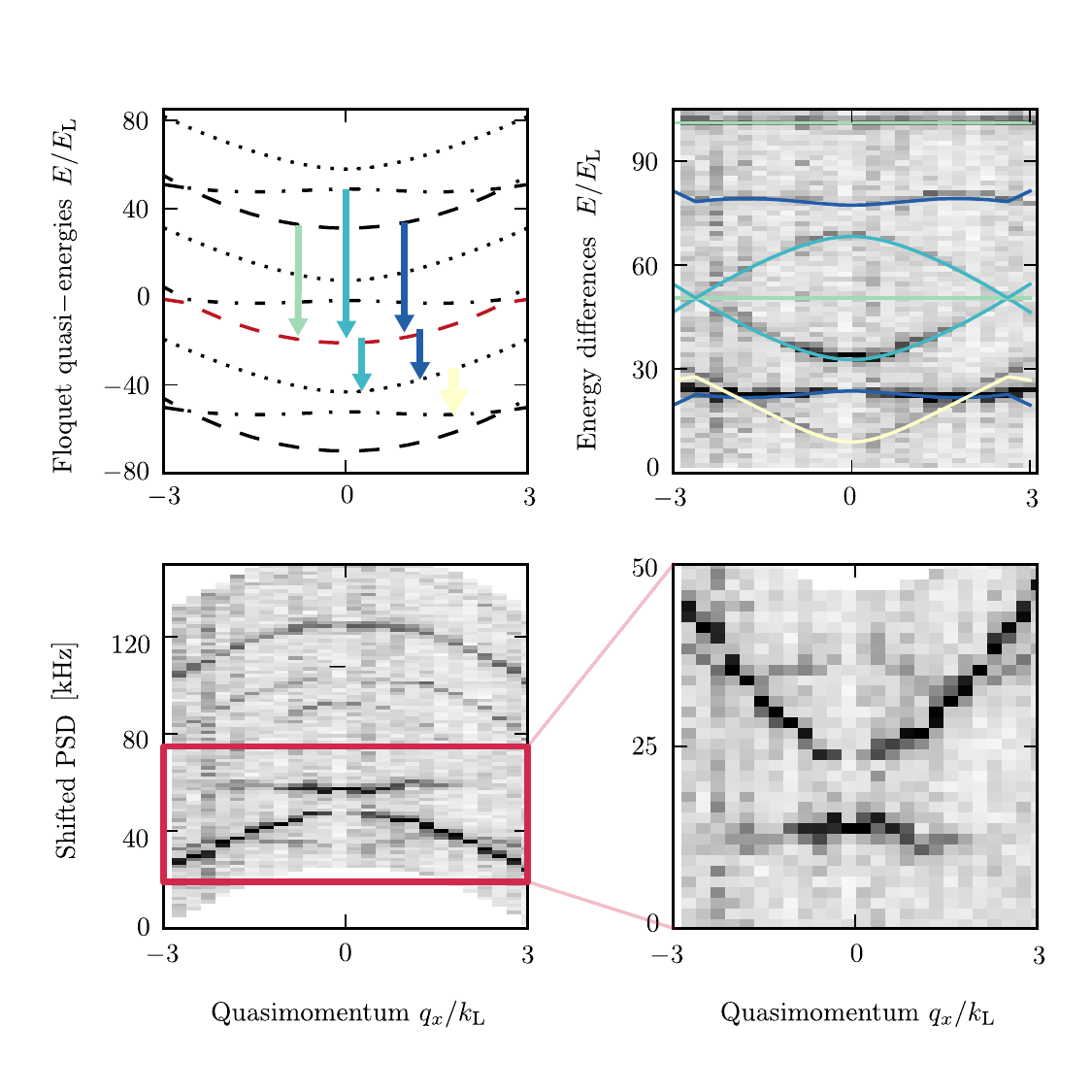}
		\caption
		{  {\bf (a)} Floquet quasi-energy spectrum of a SOC Hamiltonian with periodic coupling strength. The red line represents the eigenstate that has the largest overlap with the initial $\ket{m_F=0}$ state. The arrows indicate the energies of the states that have non-zero overlap with the initial state and can be measured with Fourier transform spectroscopy.
		{\bf (b)} PSD of the occupation probability and numerically calculated energy differences between the levels indicated by the arrows on panel (a).
		{\bf (c)} PSD shifted by a quadratic term $-\hbar^2 q^2_x/2m^*$. The red box indicates the region of interest where we can recover the SOC spectrum.
		{\bf (d)} We invert the frequency axis and shift it by $\delta\omega$.   
		}
		\label{fig:Figure7}
	\end{center}
\end{figure*}

Figure \ref{fig:Figure7} illustrates in detail the steps that we take to obtain the dispersion for the periodically driven SOC cases. The red line in panel a) represents a level within a Floquet manifold that has the largest overlap integral with the initial $\ket{m_F=0, q=0}$ state. The peaks in the PSD correspond to energy differences between the marked level and the levels in neighboring Floquet manifolds pointed by the colored arrows. We show the theoretically computed energy differences on top of the measured PSD in panel b). The lowest frequency dominant peaks of the PSD correspond to energy differences with the adjacent lower Floquet manifold. To properly recover the SOC dispersion we need to shift the PSD by a negative quadratic term $-\hbar^2q_x^2/2m^{*}$ as we show on panel c). We finally invert the frequency axis and shift it by $\delta\omega$. Including the effective mass to reconstruct the spectrum of the time-independent SOC case, amounts to shifting the PSD by a positive quadratic term.

\section*{Appendix B: Effective Hamiltonian}

To get the effective Floquet Hamiltonian $\hat{H}_{Fl}$ from the time dependent SOC Hamiltonian in equation \ref{Eq:SOCone}, we apply a transformation $\hat{U}(t)$ such that the time evolution is given by the transformed Hamiltonian  $\hat{H}'(t)=\hat{U}^{\dagger}(t)\hat{H}(t)\hat{U}(t)-\hbar\hat{U}^{\dagger}(t)\partial_t\hat{U}(t)$. We choose the transformation
\begin{equation}
\hat{U}(t)=\mathrm{exp}[-i\frac{\Omega}{\delta\omega}\sin(\delta\omega t)\hat{F}_x].
\end{equation}

$\hat{H}'(t)$ has terms proportional to $\sin(\Omega/\delta\omega\sin(\delta\omega t))$, $\sin^2(\Omega/\delta\omega\sin(\delta\omega t))$, $\cos(\Omega/\delta\omega\sin(\delta\omega t))$ and $\cos^2(\Omega/\delta\omega\sin(\delta\omega t))$ which we simplify using the Jacobi-Anger expansion
\begin{align*}
&\cos(z\sin\theta)= J_0(z) + 2\sum_{n=1}^{\infty}J_{2n}(z)\cos(2n\theta) \approx J_0(z) \\
&\sin(z\sin\theta)= 2\sum_{n=0}^{\infty}J_{2n+1}(z)\sin((2n+1)\theta) \approx 0,
\end{align*} 
to obtain the effective time independent Hamiltonian $\hat{H}_{Fl}$.
\section*{References}

\bibliographystyle{iopart-num}
\bibliography{MolmerSorensen.bib}


 



\end{document}